\documentclass[12pt]{article}
\usepackage{amssymb}

\usepackage{amsmath}
\usepackage{epsfig}

\begin{document}

\title{Dynamics and coding of a biologically-motivated network}
\author{Carlos Aguirre\thanks{%
Computer Engineering Dept., Univ. Aut\'{o}noma de Madrid, Spain}, Jo\~{a}o
Martins\thanks{%
LabSEI, EST, IPS, Estefanilha 2914-508 Set\'{u}bal, Portugal} and R. Vilela
Mendes\thanks{%
Grupo de F\'{i}sica-Matem\'{a}tica, Complexo Interdisciplinar UL, Av. Gama
Pinto 2, 1649-003 Lisboa, Portugal} \thanks{%
corresponding author, vilela@cii.fc.ul.pt}}
\date{}
\maketitle

\begin{abstract}
A four-node network consisting of a negative loop controlling a positive one
is studied. It models some of the features of the p53 gene network. Using
piecewise linear dynamics with thresholds, the allowed dynamical classes are
fully characterized and coded. The biologically relevant situations are
identified and conclusions drawn concerning the effectiveness of the p53
network as a tumour inhibitor mechanism.
\end{abstract}

\section{Introduction}

In the modeling of biological processes, dynamical networks play an
important role. Metabolic processes of living beings are networks, the nodes
being the substrates, which are linked together whenever they participate in
the same biochemical reaction. Protein-protein as well as gene expression
and regulation are also dynamical networks.

A central role in the analysis of biological networks is played by the
oriented circuits also called \textit{feedback loops }[Snoussi \& Thomas,
1993], [Thomas et al., 1995]. Feedback loops are positive or negative
according to whether they have an even or odd number of negative
interactions. The general understanding is that positive loops generate
multistability and negative loops generate homeostasis, that is, the
variables in the loop tend to middle range values, with or without
oscillations [Snoussi, 1998], [Gouz\'{e}, 1998].

An interesting situation, not much studied yet, arises when positive and
negative loops interact and control each other. Here, one such network is
studied. Besides its interest as a dynamical system, it also tries to
capture some of the features of the p53 network.

The p53 gene was one of the first tumour-suppressor genes to be identified,
its protein acting as an inhibitor of uncontrolled cell growth. The p53
protein has been found not to be acting properly in most human cancers, due
either to mutations in the gene or inactivation by viral proteins or
inhibiting interactions with other cell products (There are however a number
of cases where apparently normal p53 does not to achieve tumour control).
Apparently not required for normal growth and development, p53 is critical
in the prevention of tumour development, contributing to DNA repair,
inhibiting angiogenesis and growth of abnormal or stressed cells [May \&
May, 1999] [Vogelstein et al., 2000] [Woods \& Vousden, 2001] [Taylor et
al., 1999] [Vousden, 2000]. In addition to its beneficial anticancer
activities it may also have some detrimental effects in human aging
[Sharpless \& DePinho, 2002]

The p53 gene does not act by itself, but through a very complex network of
interactions [Kohn, 1999]. The simplified network that is discussed in this
paper, although not being accurate in biological detail, tends to capture
the essential features of the p53 network as it is known today. In
particular, several different products and biological mechanisms are lumped
together into a single node when their function is identical. The network is
depicted in Fig.1. The arrows and signs denote the excitatory or inhibitory
action of each node on the others and the letters $b,c,p,m,g$ denote their
intensities (or concentrations).

\begin{figure}[htb]
\begin{center}
\psfig{figure=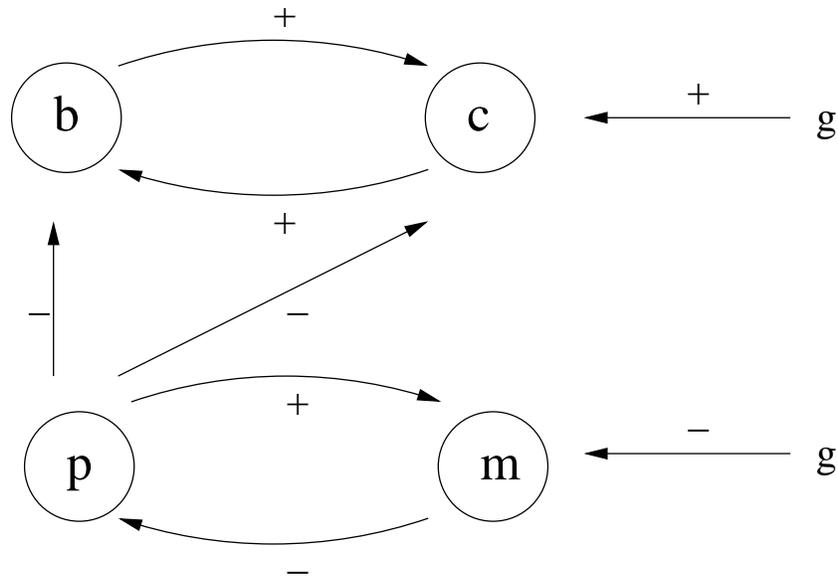,width=11truecm}
\end{center}
\caption{A network with a controlled negative loop controlling a positive
one: A simplified p53 network}
\end{figure}

The p53 protein ($p$) is assumed to be produced at a fixed rate and to be
degraded after ubiquitin labelling. The MDM2 protein being one of the main
enzymes involved in ubiquitin labelling\footnote{%
MDM2 binding to p53 can also inhibit p53's transcription}, the inhibitory
node is denoted ($m$). There is a positive feedback loop from p53 to MDM2,
because the p53 protein, binding to the regulatory region of the MDM2 gene,
stimulates the transcription of this gene into mRNA.

Under normal circumstances the network is ``off'' or operates at a low
level. The main activation pathways are the detection of cell anomalies,
like DNA damage, or abnormal growth signals, such as those resulting from
the expression of several oncogenes. They inhibit the degradation of the p53
protein, which may then reach a high level. There are several distinct
activation pathways. For example, in some cases phosphorylation of the p53
protein blocks its interaction with MDM2 and in others it is a protein that
binds to MDM2 and inhibits its action. However, the end result being a
decrease in the MDM2 efficiency, they may both be described as an inhibitory
input control ($g$) to the MDM2 node.

The p53 protein controls cell growth and proliferation ($c$), either by
blocking the cell division cycle, or activating apoptosis or inhibiting the
blood-vessel formation ($b$) that is stimulated by several tumors and is
essential for its growth. Therefore $c$ and $b$ form a positive loop.
Expression of oncogenes or abnormalities is coded as an external control ($g$%
) that operates both on the $p-m$ negative loop and the $c-b$ positive loop.
These two (controlled loops) will first be studied separately and later
their interaction is taken into account.

The dynamical evolution of the network is modeled by a piecewise-linear
discrete-time system similar to the one used in [Volchenkov \& Lima, 2003]
for other gene expression networks, namely 
\begin{equation}
\begin{array}{lll}
p\left( t+1\right) & = & a_{p}p(t)+W_{pm}H(T_{m}-m(t)) \\ 
m\left( t+1\right) & = & a_{m}m(t)+W_{mp}H(p(t)-T_{p})+W_{mg}H\left(
T_{g}-g\right) \\ 
c\left( t+1\right) & = & 
a_{c}c(t)+W_{cb}H(b(t)-T_{b})+W_{cp}H(T_{p}-p(t))+W_{cg}H(g-T_{g}) \\ 
b\left( t+1\right) & = & a_{b}b(t)+W_{bc}H(c(t)-T_{c})+W_{bp}H(T_{p}-p(t))
\end{array}
\label{1.1}
\end{equation}
$H$ is the Heaviside function, the $T_{i}$'s are the thresholds and in all
cases 
\begin{equation}
a_{i}+\sum_{k}W_{ik}=1  \label{1.2}
\end{equation}
$W_{ik}>0$. This condition insures that all intensities remain in interval [$%
0,1$].

A similar network was studied in [Vilela Mendes, 2003] modelled by a
differential equation system. However, the piecewise-linear discrete-time
model considered here allows for a more rigorous characterization of the
dynamical features of the network.

The study that we perform analyses the two basic network modules, namely the
positive and negative loops. This type of modular analysis has been proposed
before [Thieffry \& Romero, 1999] as a tool to study biological networks.
However this is an adequate methodology only if, in each case, the effect of
the other modules is taken into account as an external control. This changes
the usual conditions for multistationarity and stable periodicity. We will
follow this extended modular approach, by considering the $p-m$ loop as
controlled by the $g$ signal and the $c-b$ loop as doubly controlled by $g$
and $p$.

\section{The p-m system: a controlled negative loop}

\begin{figure}[htb]
\begin{center}
\psfig{figure=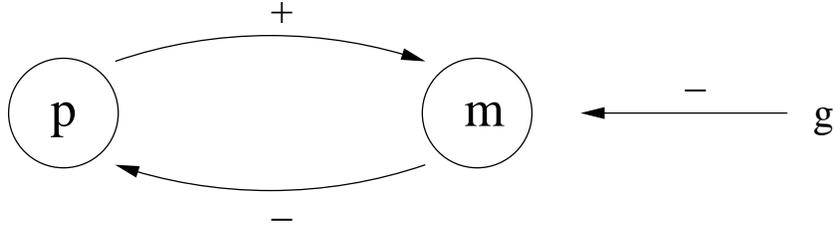,width=11truecm}
\end{center}
\caption{The $p-m$ system. A controlled negative loop}
\end{figure}

\begin{eqnarray}
p(t+1) &=&a_{p}p(t)+W_{pm}H(T_{m}-m(t))  \label{2.1} \\
m(t+1) &=&a_{m}m(t)+W_{mp}H(p(t)-T_{p})+W_{mg}H\left( T_{g}-g\right) 
\nonumber
\end{eqnarray}

\begin{figure}[htb]
\begin{center}
\psfig{figure=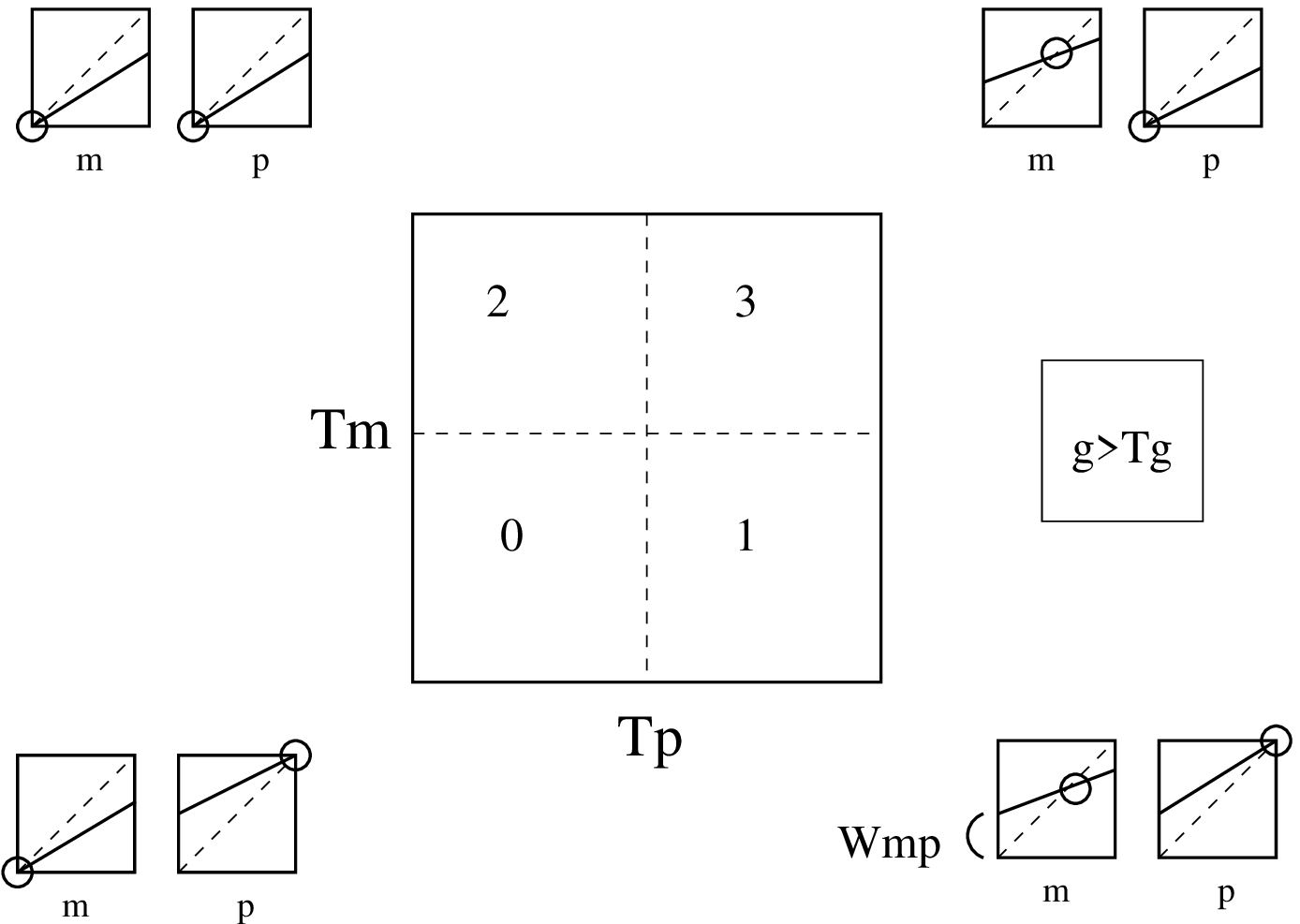,width=11truecm}
\end{center}
\caption{The $p-m$ system. The dynamical laws in each region for $g>T_{g}$}
\end{figure}

The thresholds $T_{p}$ and $T_{m}$ divide the unit square $[0,1]^{2}$ into
four distinct regions with different dynamical laws. This is illustrated in
Fig.3 (for the case $g>T_{g}$), where close to the corner of each region we
have drawn the next-time map corresponding to the evolution of $p$ and $m$.
Because the $a_{i}$'s are smaller than one, the dynamics tends to drive the
system to a fixed point. However if the fixed point of the dynamics is
outside that region, the dynamical law will change whenever the system moves
to another region. Therefore the system may never reach a fixed point and
evolve forever. This is illustrated in the Fig.4, where the regions and the
corresponding fixed points are color coded for the two cases ($g=0$ or $g=1$%
). Notice also that, for future reference, the four regions are coded as 0,
1, 2 and 3.

\begin{figure}[tbh]
\begin{center}
\psfig{figure=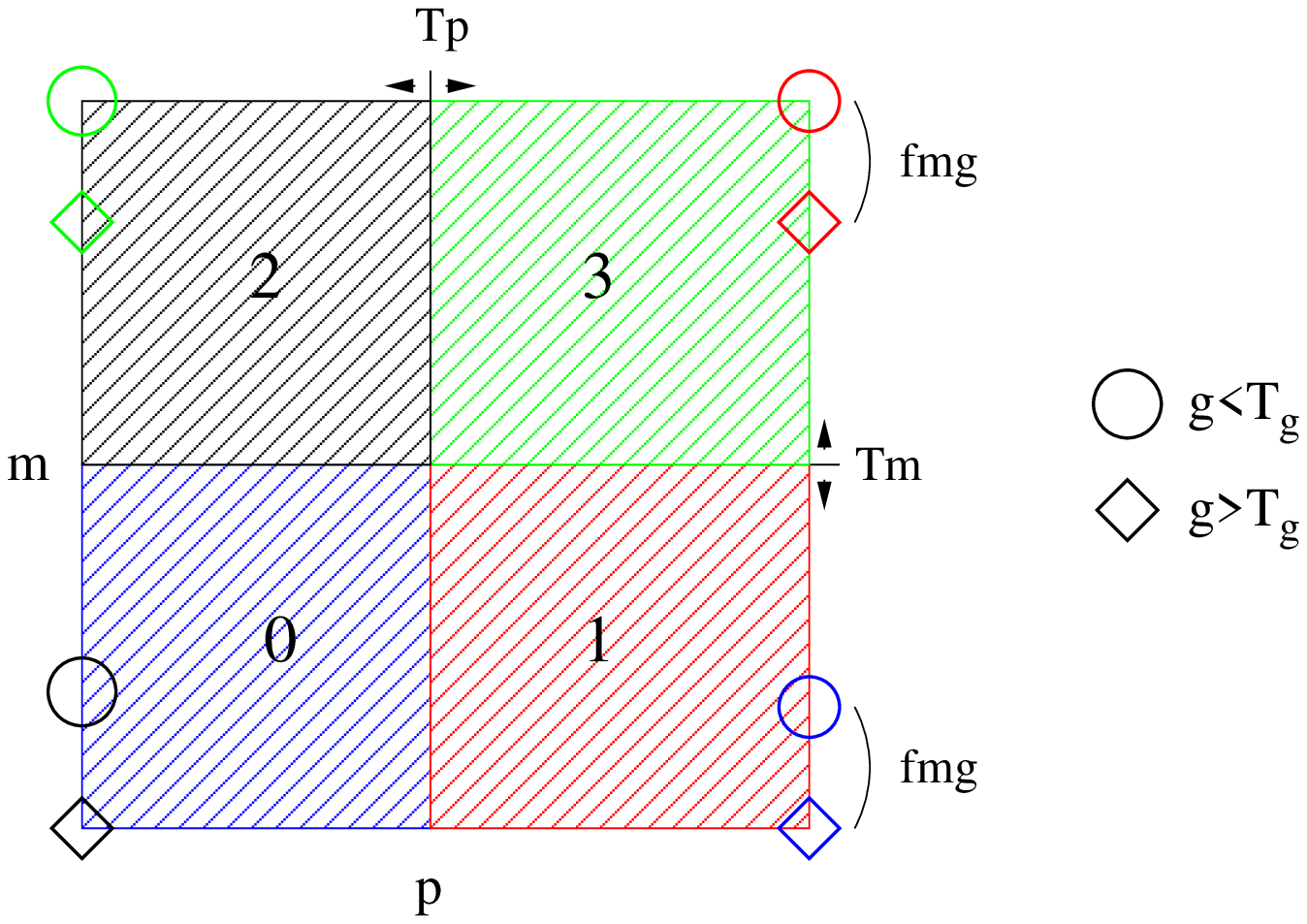,width=11truecm}
\end{center}
\caption{Colour coded dynamical regions and the corresponding fixed points
for the $p-m$ system}
\end{figure}

Define the quantities 
\begin{eqnarray}
fmp &=&\frac{W_{mp}}{1-a_{m}}  \label{2.2} \\
fmg &=&\frac{W_{mg}}{1-a_{m}}  \nonumber
\end{eqnarray}
Notice that from (\ref{1.2}) it follows 
\begin{equation}
fmp+fmg=1  \label{2.3}
\end{equation}
Examining Fig.4 and considering all qualitatively different positions of the
thresholds one obtains a complete classification of all possible dynamical
behaviors of this system, which are listed in the following table:

TABLE 1. 
\begin{equation}
\begin{tabular}{|c|c|c|c|}
\hline
& $g$ & Asympt. behavior & Coding \\ \hline
$
\begin{array}{l}
T_{m}<fmg \\ 
T_{m}>fmp
\end{array}
$ & 
\begin{tabular}{c}
$g<T_{g}$ \\ 
\\ 
--------- \\ 
\\ 
$g>T_{g}$%
\end{tabular}
& 
\begin{tabular}{c}
$p\rightarrow 0,m\rightarrow fmg$ \\ 
\\ 
------------------------ \\ 
\\ 
$p\rightarrow 1,m\rightarrow fmp$%
\end{tabular}
& 
\begin{tabular}{c}
$0\rightarrow 3\rightarrow 2$ \\ 
$\stackrel{\uparrow }{1}$ \\ 
--------------- \\ 
$2\rightarrow 0\rightarrow 1$ \\ 
$\stackrel{\uparrow }{3}$%
\end{tabular}
\\ \hline
$
\begin{array}{l}
T_{m}<fmg \\ 
T_{m}<fmp
\end{array}
$ & 
\begin{tabular}{c}
$g<T_{g}$ \\ 
\\ 
--------- \\ 
\\ 
$g>T_{g}$%
\end{tabular}
& 
\begin{tabular}{c}
$p\rightarrow 0,m\rightarrow fmg$ \\ 
\\ 
------------------------ \\ 
\\ 
oscillation
\end{tabular}
& 
\begin{tabular}{c}
$0\rightarrow 3\rightarrow 2$ \\ 
$\stackrel{\uparrow }{1}$ \\ 
--------------- \\ 
0$\rightarrow $1 \\ 
$\stackrel{\uparrow }{2}\leftarrow \stackrel{\downarrow }{3}$%
\end{tabular}
\\ \hline
$
\begin{array}{l}
T_{m}>fmg \\ 
T_{m}>fmp
\end{array}
$ & 
\begin{tabular}{c}
$g<T_{g}$ \\ 
\\ 
--------- \\ 
\\ 
$g>T_{g}$%
\end{tabular}
& 
\begin{tabular}{c}
oscillation \\ 
\\ 
------------------------ \\ 
\\ 
$p\rightarrow 1,m\rightarrow fmp$%
\end{tabular}
& 
\begin{tabular}{c}
0$\rightarrow $1 \\ 
$\stackrel{\uparrow }{2}\leftarrow \stackrel{\downarrow }{3}$ \\ 
--------------- \\ 
$2\rightarrow 0\rightarrow 1$ \\ 
$\stackrel{\uparrow }{3}$%
\end{tabular}
\\ \hline
$
\begin{array}{l}
T_{m}>fmg \\ 
T_{m}<fmp
\end{array}
$ & 
\begin{tabular}{c}
$g<T_{g}$ \\ 
\\ 
--------- \\ 
\\ 
$g>T_{g}$%
\end{tabular}
& 
\begin{tabular}{c}
oscillation \\ 
\\ 
------------------------ \\ 
\\ 
oscillation
\end{tabular}
& 
\begin{tabular}{c}
0$\rightarrow $1 \\ 
$\stackrel{\uparrow }{2}\leftarrow \stackrel{\downarrow }{3}$ \\ 
---------------- \\ 
0$\rightarrow $1 \\ 
$\stackrel{\uparrow }{2}\leftarrow \stackrel{\downarrow }{3}$%
\end{tabular}
\\ \hline
\end{tabular}
\label{2.4}
\end{equation}

The period of the oscillations will depend on the specific values of the
parameters, but the qualitative nature of the dynamics only depends on the
relative positions of the thresholds and the quantities $fmg$ and $fmp$. The
coding represents the transitions between the regions defined by the
thresholds. Depending on the numerical values of the parameters, the system
may stay for a number of time steps in each region but then it is forced to
follow the arrows in the coding scheme.

Of the above four working regimens, the first two ($T_{m}<fmg$) are the
biologically relevant ones, because in the absence of oncogenes ($g$),
normal p53 is expected to be at a low level. The other two regimens
correspond to biologically detrimental action of this gene, provoking
premature aging.

\section{The c-b system: a doubly controlled positive loop}

\begin{figure}[htb]
\begin{center}
\psfig{figure=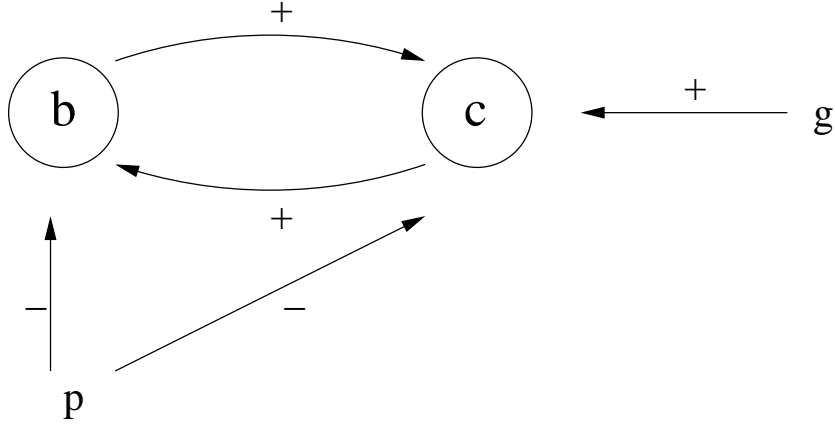,width=11truecm}
\end{center}
\caption{The c-b system: a doubly controlled positive loop}
\end{figure}

\begin{eqnarray}
b(t+1) &=&a_{b}b(t)+W_{bc}H(c(t)-T_{c})+W_{bp}H(T_{p}-p(t))  \label{3.1} \\
c(t+1) &=&a_{c}c(t)+W_{cb}H(b(t)-T_{b})+W_{cp}H(T_{p}-p(t))+W_{cg}H(g-T_{g})
\nonumber
\end{eqnarray}

Here we consider this system as a positive loop, with dynamical variables $b$
and $c$ and two controls $p$ and $g$. The relevant quantities are 
\begin{eqnarray}
fcp &=&\frac{W_{cp}}{1-a_{c}}  \label{3.2} \\
fcb &=&\frac{W_{cb}}{1-a_{c}}  \nonumber \\
fcg &=&\frac{W_{cg}}{1-a_{c}}  \nonumber
\end{eqnarray}
with 
\begin{equation}
fcp+fcb+fcg=1  \label{3.3}
\end{equation}
and 
\begin{eqnarray}
fbc &=&\frac{W_{bc}}{1-a_{b}}  \label{3.4} \\
fbp &=&\frac{W_{bp}}{1-a_{b}}  \nonumber
\end{eqnarray}
with 
\begin{equation}
fbc+fbp=1  \label{3.5}
\end{equation}
As before, the dynamics is characterized by analyzing the position of the
fixed points for the 4 possible situations ($p\lessgtr T_{p},g\lessgtr T_{g}$%
). This is illustrated in Fig.5, for particular values of the thresholds.
The regions defined by the thresholds are, in this case, coded as a, b, c,
d. We recall that the coding, listed in the tables, represents the required
dynamical transitions between the regions. Depending on the numerical values
of the parameters, the system may remain in each region a certain number of
time steps before each transition. By adjusting the parameters we may
therefore model different action delays in the network. The most important
parameters controlling the delays are the dissipation parameters $a_{i}$.

\begin{figure}[htb]
\begin{center}
\psfig{figure=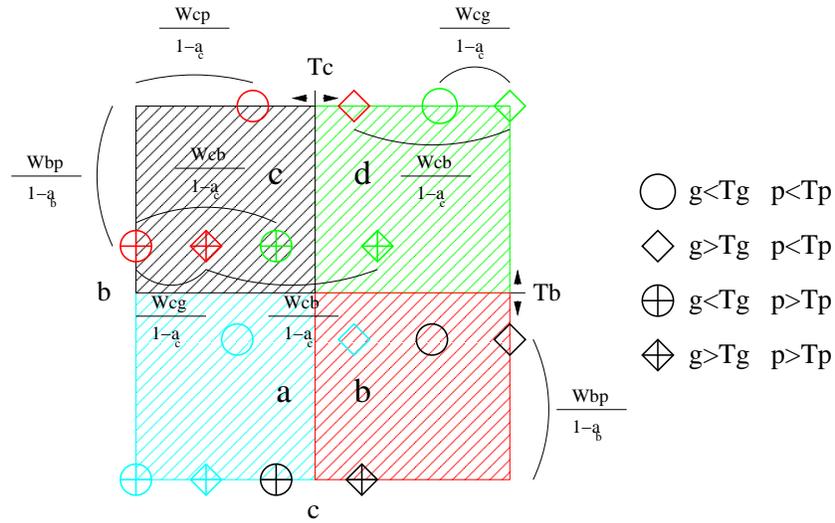,width=11truecm}
\end{center}
\caption{Colour coded dynamical regions and the corresponding fixed points
for the $c-b$ system}
\end{figure}

TABLE\ 2A. (Case $g<T_{g}$ and $p<T_{p}$) 
\[
\begin{tabular}{|c|c|c|}
\hline
$g<T_{g}$ , $p<T_{p}$ & Asympt. behavior & Coding \\ \hline
$
\begin{array}{l}
T_{c}<fcp \\ 
T_{b}<fbp
\end{array}
$ & $
\begin{array}{l}
b\rightarrow 1 \\ 
c\rightarrow fcp+fcb
\end{array}
$ & 
\begin{tabular}{c}
a$\rightarrow $d$\leftarrow $b \\ 
$\stackrel{\uparrow }{\text{c}}$%
\end{tabular}
\\ \hline
$
\begin{array}{l}
T_{c}<fcp \\ 
T_{b}>fbp
\end{array}
$ & $
\begin{array}{l}
b\rightarrow 1 \\ 
c\rightarrow fcp+fcb
\end{array}
$ & 
\begin{tabular}{c}
a$\rightarrow $b$\rightarrow $d \\ 
$\stackrel{\uparrow }{\text{c}}$%
\end{tabular}
\\ \hline
$
\begin{array}{l}
fcp+fcb>T_{c}>fcp \\ 
T_{b}<fbp
\end{array}
$ & $
\begin{array}{l}
b\rightarrow 1 \\ 
c\rightarrow fcp+fcb
\end{array}
$ & 
\begin{tabular}{c}
a$\rightarrow $c$\rightarrow $d \\ 
$\stackrel{\uparrow }{\text{b}}$%
\end{tabular}
\\ \hline
$
\begin{array}{l}
fcp+fcb>T_{c}>fcp \\ 
T_{b}>fbp
\end{array}
$ & $
\begin{array}{l}
b\rightarrow 1,c\rightarrow fcp+fcb \\ 
\text{or} \\ 
b\rightarrow fbp,c\rightarrow fcp \\ 
\text{or} \\ 
\text{oscillation}
\end{array}
$ & 
\begin{tabular}{c}
d$\circlearrowleft $ \\ 
\\ 
a$\circlearrowleft $ \\ 
\\ 
b$\leftrightarrows $c
\end{tabular}
\\ \hline
$
\begin{array}{l}
T_{c}>fcp+fcb \\ 
T_{b}<fbp
\end{array}
$ & $
\begin{array}{l}
b\rightarrow fbp \\ 
c\rightarrow fcp+fcb
\end{array}
$ & 
\begin{tabular}{c}
a$\rightarrow $c$\leftarrow $b \\ 
$\stackrel{\uparrow }{\text{d}}$%
\end{tabular}
\\ \hline
$
\begin{array}{l}
T_{c}>fcp+fcb \\ 
T_{b}>fbp
\end{array}
$ & $
\begin{array}{l}
b\rightarrow fbp \\ 
c\rightarrow fcp
\end{array}
$ & 
\begin{tabular}{c}
b$\rightarrow $c$\rightarrow $a \\ 
$\stackrel{\uparrow }{\text{d}}$%
\end{tabular}
\\ \hline
\end{tabular}
\]
TABLE\ 2B. (Case $g<T_{g}$ and $p>T_{p}$) 
\[
\begin{tabular}{|c|c|c|}
\hline
$g<T_{g}$ , $p>T_{p}$ & Asympt. behavior & Coding \\ \hline
$
\begin{array}{l}
T_{c}<fcb \\ 
T_{b}<fbc
\end{array}
$ & $
\begin{array}{l}
b\rightarrow 0,c\rightarrow 0 \\ 
\text{or} \\ 
b\rightarrow fbc,c\rightarrow fcb \\ 
\text{or} \\ 
\text{oscillation}
\end{array}
$ & 
\begin{tabular}{c}
a$\circlearrowleft $ \\ 
\\ 
d$\circlearrowleft $ \\ 
\\ 
b$\leftrightarrows $c
\end{tabular}
\\ \hline
$
\begin{array}{l}
T_{c}<fcb \\ 
T_{b}>fbc
\end{array}
$ & $
\begin{array}{l}
b\rightarrow 0 \\ 
c\rightarrow 0
\end{array}
$ & 
\begin{tabular}{c}
c$\rightarrow $b$\rightarrow $a \\ 
$\stackrel{\uparrow }{\text{d}}$%
\end{tabular}
\\ \hline
$
\begin{array}{l}
T_{c}>fcb \\ 
T_{b}<fbc
\end{array}
$ & $
\begin{array}{l}
b\rightarrow 0 \\ 
c\rightarrow 0
\end{array}
$ & 
\begin{tabular}{c}
b$\rightarrow $c$\rightarrow $a \\ 
$\stackrel{\uparrow }{\text{d}}$%
\end{tabular}
\\ \hline
$
\begin{array}{l}
T_{c}>fcb \\ 
T_{b}>fbc
\end{array}
$ & $
\begin{array}{l}
b\rightarrow 0 \\ 
c\rightarrow 0
\end{array}
$ & 
\begin{tabular}{c}
c$\rightarrow $a$\leftarrow $b \\ 
$\stackrel{\uparrow }{\text{d}}$%
\end{tabular}
\\ \hline
\end{tabular}
\]
TABLE 2C. (Case $g>T_{g}$ and $p<T_{p}$) 
\[
\begin{tabular}{|c|c|c|}
\hline
$g>T_{g}$ , $p<T_{p}$ & Asympt. behavior & Coding \\ \hline
$
\begin{array}{l}
T_{c}<fcg+fcp \\ 
T_{b}<fbp
\end{array}
$ & $
\begin{array}{l}
b\rightarrow 1 \\ 
c\rightarrow 1
\end{array}
$ & 
\begin{tabular}{c}
c$\rightarrow $d$\leftarrow $b \\ 
$\stackrel{\uparrow }{\text{a}}$%
\end{tabular}
\\ \hline
$
\begin{array}{l}
T_{c}<fcg+fcp \\ 
T_{b}>fbp
\end{array}
$ & $
\begin{array}{l}
b\rightarrow 1 \\ 
c\rightarrow 1
\end{array}
$ & 
\begin{tabular}{c}
a$\rightarrow $b$\rightarrow $d \\ 
$\stackrel{\uparrow }{\text{c}}$%
\end{tabular}
\\ \hline
$
\begin{array}{l}
T_{c}>fcg+fcp \\ 
T_{b}<fbp
\end{array}
$ & $
\begin{array}{l}
b\rightarrow 1 \\ 
c\rightarrow 1
\end{array}
$ & 
\begin{tabular}{c}
a$\rightarrow $c$\rightarrow $d \\ 
$\stackrel{\uparrow }{\text{b}}$%
\end{tabular}
\\ \hline
$
\begin{array}{l}
T_{c}>fcg+fcp \\ 
T_{b}>fbp
\end{array}
$ & $
\begin{array}{l}
b\rightarrow 1,c\rightarrow 1 \\ 
\text{or} \\ 
b\rightarrow fbp,c\rightarrow fcg+fcp \\ 
\text{or} \\ 
\text{oscillation}
\end{array}
$ & 
\begin{tabular}{c}
d$\circlearrowleft $ \\ 
\\ 
a$\circlearrowleft $ \\ 
\\ 
b$\leftrightarrows $c
\end{tabular}
\\ \hline
\end{tabular}
\]
TABLE 2D. (Case $g>T_{g}$ and $p>T_{p}$) 
\[
\begin{tabular}{|c|c|c|}
\hline
$g>T_{g}$ , $p>T_{p}$ & Asympt. behavior & Coding \\ \hline
$
\begin{array}{l}
T_{c}<fcg \\ 
T_{b}<fbc
\end{array}
$ & $
\begin{array}{l}
b\rightarrow fbc \\ 
c\rightarrow fcg+fcb
\end{array}
$ & 
\begin{tabular}{c}
a$\rightarrow $b$\rightarrow $d \\ 
$\stackrel{\uparrow }{\text{c}}$%
\end{tabular}
\\ \hline
$
\begin{array}{l}
T_{c}<fcg \\ 
T_{b}>fbc
\end{array}
$ & $
\begin{array}{l}
b\rightarrow fbc \\ 
c\rightarrow fcg
\end{array}
$ & 
\begin{tabular}{c}
a$\rightarrow $b$\leftarrow $d \\ 
$\stackrel{\uparrow }{\text{c}}$%
\end{tabular}
\\ \hline
$
\begin{array}{l}
fcg+fcb>T_{c}>fcg \\ 
T_{b}<fbc
\end{array}
$ & $
\begin{array}{l}
b\rightarrow fbc,c\rightarrow fcg+fcb \\ 
\text{or} \\ 
b\rightarrow 0,c\rightarrow fcg \\ 
\text{or} \\ 
\text{oscillation}
\end{array}
$ & 
\begin{tabular}{c}
d$\circlearrowleft $ \\ 
\\ 
a$\circlearrowleft $ \\ 
\\ 
b$\leftrightarrows $c
\end{tabular}
\\ \hline
$
\begin{array}{l}
fcg+fcb>T_{c}>fcg \\ 
T_{b}>fbc
\end{array}
$ & $
\begin{array}{l}
b\rightarrow 0 \\ 
c\rightarrow fcg
\end{array}
$ & 
\begin{tabular}{c}
c$\rightarrow $b$\rightarrow $a \\ 
$\stackrel{\uparrow }{\text{d}}$%
\end{tabular}
\\ \hline
$
\begin{array}{l}
T_{c}>fcg+fcb \\ 
T_{b}<fbc
\end{array}
$ & $
\begin{array}{l}
b\rightarrow 0 \\ 
c\rightarrow fcg
\end{array}
$ & 
\begin{tabular}{c}
b$\rightarrow $c$\rightarrow $a \\ 
$\stackrel{\uparrow }{\text{d}}$%
\end{tabular}
\\ \hline
$
\begin{array}{l}
T_{c}>fcg+fcb \\ 
T_{b}>fbc
\end{array}
$ & $
\begin{array}{l}
b\rightarrow 0 \\ 
c\rightarrow fcg
\end{array}
$ & 
\begin{tabular}{c}
c$\rightarrow $a$\leftarrow $b \\ 
$\stackrel{\uparrow }{\text{d}}$%
\end{tabular}
\\ \hline
\end{tabular}
\]
\newline
Notice that the oscillations coded as b$\leftrightarrows $c can only be
realized if the dynamics jumps directly between these two regions.
Otherwise, if in the course of the dynamical evolution, the system crosses
the regions a or d, it will be attracted to their fixed points. In this
sense this oscillation code is not as robust as the one ( 
\begin{tabular}{c}
0$\rightarrow $1 \\ 
$\stackrel{\uparrow }{2}\leftarrow \stackrel{\downarrow }{3}$%
\end{tabular}
) in the $p-m$ system and might only be implemented for exceptional
parameter values.

\section{Biological implications and the coupled system}

The analysis performed in Sections 2 and 3 is quite general and applies to
any parameter region. However, not all parameter values will model
biologically normal conditions. For example, in the absence of cell
abnormalities or oncogenes ($g<T_{g}$), normal p53 will be at a low level.
Therefore, as follows from Table 1. 
\begin{equation}
T_{m}<fmg  \label{4.1}
\end{equation}
On the other hand, on the absence of oncogenes and low p53, that is, under
non-pathological conditions, cell growth should not be explosive. Therefore
this condition requires for the model 
\begin{equation}
\begin{array}{l}
T_{c}>fcp+fcb \\ 
T_{b}>fbp
\end{array}
\label{4.2}
\end{equation}
as follows from Table 2A.

With the above (normal condition) requirements we are now ready to analyze
the behavior for the other regimens. First, it follows from (\ref{4.2}) that
also $T_{c}>fcb$. Therefore one sees, from table 2B that if $g<T_{g}$ and $%
p>T_{p}$, then $b\rightarrow 0$ and $c\rightarrow 0$. That is, in the
absence of oncogenes, expression of p53 completely blocks cell growth.

On the other hand, one notices, from table 2C, that in all cases, when $%
g>T_{g}$ (presence of oncogenes) and $p<T_{p}$, there is explosive cell
growth ($c\rightarrow 1$). Only in the special case $T_{c}>fcg+fcp$ and $%
T_{b}>fbp$ there are other solutions, but even in this case $c\rightarrow 1$
is a possible solution.

Finally, the most interesting situation is to analyze the effect of
expressed p53 ($p>T_{p}$) in the presence of oncogenes ($g>T_{g}$) - Table
2D :

- For $T_{c}>fcg+fcb$ (and any $T_{b}$) p53 effectively controls cell
growth, keeping it at a low level ($c\rightarrow fcg$). In this network
model, a large value of $T_{c}$ is related to the effectiveness of cell
growth to stimulate the required increased level of blood supply. The larger 
$T_{c}$ the less effective is the stimulation signal.

- For $T_{c}$ values smaller than $fcg+fcb$ the inhibitory effect of p53 may
not be so effective. For example, for $fcg+fcb>T_{c}>fcg$ and $T_{b}<fbc$
there are several solutions and the outcome will depend on the initial
conditions. For $T_{c}<fcg$ and $T_{b}<fbc$ no control is possible.

For parameter values of $T_{m}$ and $T_{p}$ for which the $p-m$ system is
driven towards a fixed point, the behavior of the coupled system may be read
directly from the $c-b$ tables of the previous section and the conclusions
are as listed above. For oscillatory conditions on the $p-m$ system, the
behavior is more complex and several different regimens may be accessed.
Here we will illustrate, by direct numerical simulation, the behavior of the
coupled system, for regions of parameter values which, as discussed above,
are biologically relevant ($T_{m}<fmg,T_{c}>fcp+fcb$ and $T_{b}>fbp$).

\begin{figure}[htb]
\begin{center}
\psfig{figure=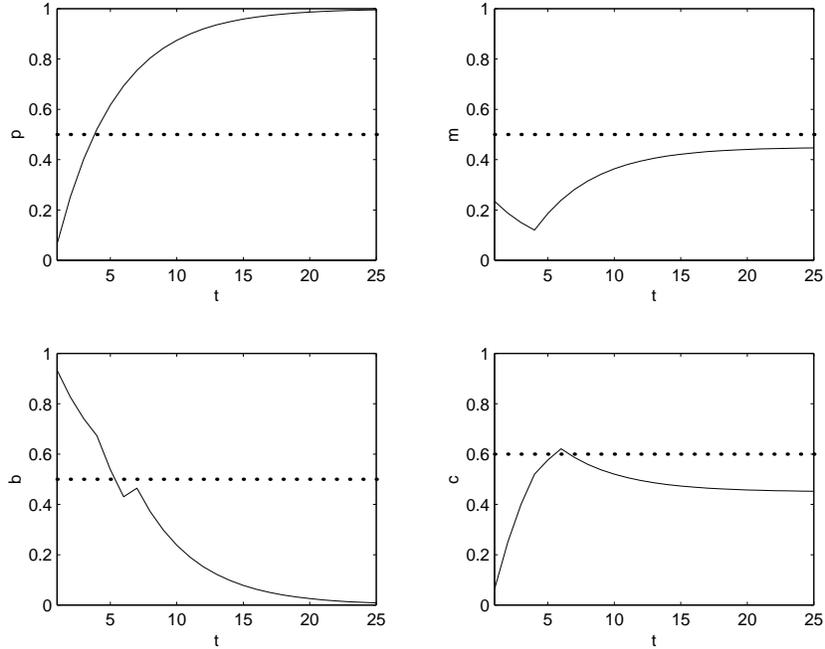,width=11truecm}
\end{center}
\caption{Dynamical evolution of $p$, $m$, $b$, and $c$ for $%
fmg=0.55,\;fbp=0.4,\;fcb=0.35,\;fcp=0.2,\;a_{i}=0.8,\;T_{p}=0.5,\;T_{m}=0.5,%
\;T_{c}=0.6,\;T_{b}=0.5,\;g>T_{g}$}
\end{figure}

\begin{figure}[htb]
\begin{center}
\psfig{figure=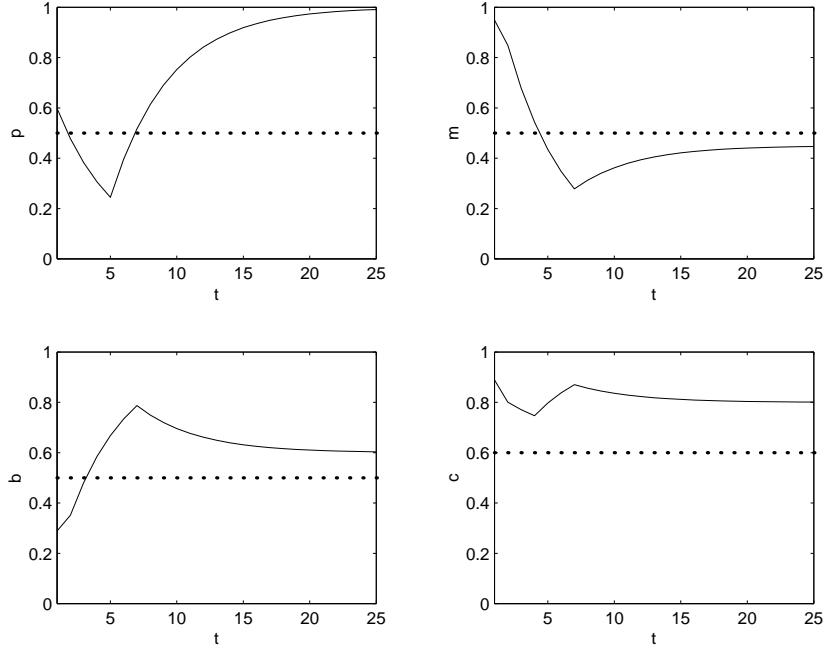,width=11truecm}
\end{center}
\caption{Same as Fig.7 with different initial conditions}
\end{figure}

Figs.7 and 8 show the dynamical evolution of the variables $p$, $m$, $b$,
and $c$ for $g>T_{g}$ and the parameter values 
\[
fmg=0.55,\;fbp=0.4,\;fcb=0.35,\;fcp=0.2 
\]
and 
\[
T_{p}=0.5,\;T_{m}=0.5,\;T_{c}=0.6,\;T_{b}=0.5 
\]
This corresponds to the situation 
\[
T_{m}<fmg,\;T_{m}>fmp 
\]
in the $p-m$ system and 
\[
fcg+fcb>T_{c}>fcg,\;T_{b}<fbc 
\]
in the $c-b$ system.

The parameter values are the same for both figures. One sees that depending
on the initial conditions, in one case the p53 action keeps $c$ at a low
level and in the other it does not.

\begin{figure}[htb]
\begin{center}
\psfig{figure=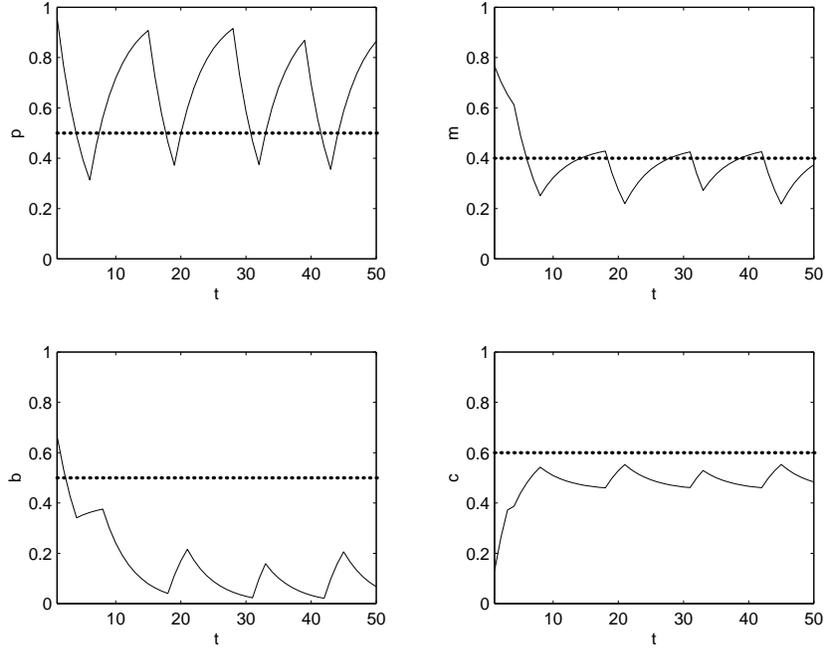,width=11truecm}
\end{center}
\caption{Same as Fig.7 with $T_{m}=0.4$ instead of $T_{m}=0.5$}
\end{figure}

\begin{figure}[htb]
\begin{center}
\psfig{figure=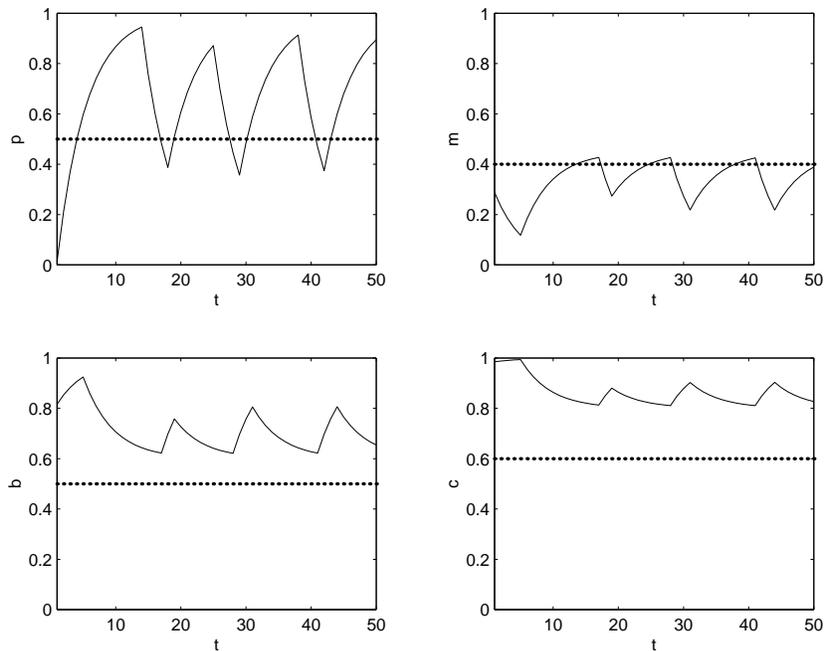,width=11truecm}
\end{center}
\caption{Same as Fig.9 with different initial conditions}
\end{figure}

Figs. 9 and 10 use the same parameters for the $c-b$ system, the only change
being the value of $T_{m}$ that now is $0.4$, that is, $T_{m}<fmg$ and $%
T_{m}<fmp$. In this case the $p-m$ system oscillates instead of converging
to a fixed point, but the action on the $c-b$ system is similar and it again
depends on the initial conditions.

\section{Remarks and conclusions}

1 - A simplified four-node model displays biologically reasonable features
which might, at least, serve as a toy model for the p53 network. The model
allows for a complete mathematical characterization of its dynamical
solutions and how they depend on the parameter values (thresholds and
couplings). In biology, to change the threshold positions corresponds to
stimulate or inhibit the mechanisms of gene expression. Therefore the
dynamical identification of their role on the dynamics of the network might
be a guide for therapeutic action.

2 - For very large values of the $T_{c}$ threshold, our model p53 ($p$)
achieves control of cell growth in the presence of oncogenes. However, as
illustrated before, there is a biologically reasonable region of parameters
for which the successful action of $p$, as an inhibitor of cell growth,
strongly depends on the initial conditions. In practice it means that it is
only effective if the $p$ level is already large enough before the positive
feedback effect of $c$ on $b$ starts to play a role.

It is known that in about half of human cancers p53 action is lost through
mutation in the p53 gene. In many of the remaining tumors, the p53 gene is
intact but it does not achieve the proper response [Woods \& Vousden, 2001].
Several mechanisms have been proposed to explain the loss of genetically
normal p53 action. They include abnormal conformation , targeting for
degradation by viral proteins, defective localization to the nucleus,
amplification of MDM2, loss of ability to inhibit MDM2 through proteins such
as p14$^{ARF}$ or loss of kinases that phosphorylate MDM2 and p53. What our
simple model suggests is that there is a range of biological parameters for
which success or failure of the p53 action is very much a question of
chance, that is, it depends on the initial conditions of the process. This
would be a purely dynamic explanation for the failure of normal p53.

3 - The actual biological p53 network is highly complex and contains very
many nodes and pathways [Kohn, 1999]. Our toy model has concentrated on
extracting the abstract functional actions rather than reproducing the
biological detail. However, there are still many other dynamical mechanisms
in the organic control of DNA abnormalities or pathological cell
proliferation. A number of other genes either play a similar role or
interact with p53. On the other hand some of the ``defense'' mechanisms of
cancer cells are directly related to the operation of the basic network. For
example, oxygen starving (hypoxia) of tumor cells by lack of blood supply
(the $b$ node intensity) activates the CXCR4 gene and this activation causes
the tumor cells to migrate to other organs (metastasis) [Staller et al.,
2003]. Inclusion of this and other qualitatively different effects seems an
interesting possibility.

4 - Sections 2 and 3 contain a complete classification of the fixed versus periodic
nature of the orbits. However, as mentioned before, the actual period of the 
oscillations depends strongly on the values of the parameters. In Fig.11
we illustrate this dependence by color coding the logarithm of the period of the orbits,
 in the negative loop, as a function of the parameters $a_{m}$ and $a_{p}$.
The parameters are chosen such that $W_{pm}=1-a_{m}$, $W_{mp}=0.5(1-a_{m})$ and
$W_{mg}=0.5(1-a_{m})$. these conditions satisfy the oscillation
condition and mantain normalization.

\begin{figure}[htb]
\begin{center}
\psfig{figure=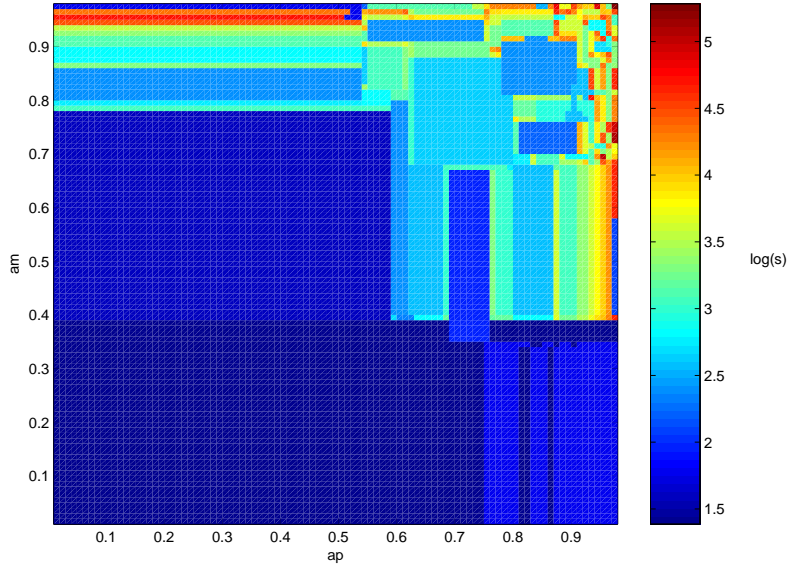,width=11truecm}
\end{center}
\caption{Period of the closed orbits in the negative loop. $g>T_{g}$, $T_{p}=0.5$,
$T_{m}=0.4$, $fmg=0.5$}
\end{figure}

To visualize the orbits corresponding to any parameter value a tool 
for Windows is avaiable for download at 
\tt{http://www.ii.uam.es/~aguirre/p53\_emu.zip}.

The behavior of the periods is rather discontinuous, switching from
small to large values for small parameter changes. This occurs,
inparticular, when some points of the orbit lie close to the thresholds.

In Fig.12 the rotation number $i$ (the number of cycles divided 
by the period of the orbit, is depicted. This quantity displays
a more continous behavior that the period of the orbits. 
Highest values of $i$ correspond to high values of both $a_{m}$ 
and $a_{p}$. However, there are still some discontinuities due 
mainly to orbits close to the intersection of the thresholds.

\begin{figure}[htb]
\begin{center}
\psfig{figure=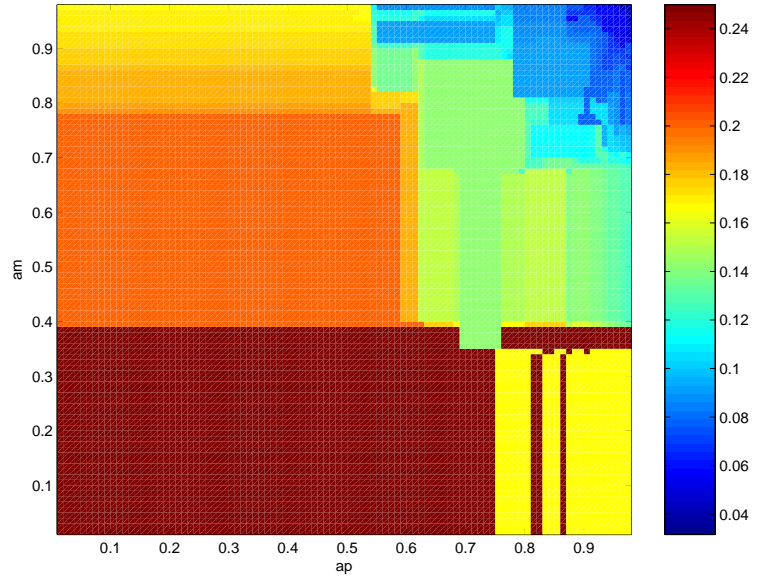,width=11truecm}
\end{center}
\caption{Rotation number of the periodic orbits in the negative 
loop. The parameters are the same as Fig.11}
\end{figure}

{\LARGE References}

Gouz\'{e}, J.-L. [1998]; \textit{Positive and negative circuits in dynamical
systems}, J. of Bio. Syst. 6, 11-15.

Kohn, K. W. [1999]; \textit{Molecular interaction map of the mammalian cell
cycle control and DNA repair systems}, Molecular Biol. of the Cell 10,
2703-2734.

May, P. \& May, E. [1999]; \textit{Twenty years of p53 research: Structural
and functional aspects of the p53 protein}, Oncogene 18, 7621-7636.

Sharpless, N. E. \& DePinho, R. A. [2002]; \textit{p53: Good cop/Bad cop},
Cell 110, 9-12.

Snoussi, E. H. [1998]; \textit{Necessary conditions for multistationarity
and stable periodicity}, J. of Bio. Syst. 6, 3-9.

Snoussi, E. H. \& Thomas, R. [1993]; \textit{Logical identification of all
steady states: The concept of feedback loop-characteristic state}, Bull.
Math. Biology 55, 973-991.

Staller, P. et al. [2003]; \textit{Chemokine receptor CXCR4 downregulated by
von Hippel-Lindau tumour suppressor pVHL}, Nature 425, 307-311.

Taylor, W. R., DePrimo, S. E., Agarwal, A., Agarwal, M. L., Sch\"{o}ntal, A.
H., Katula, K. S. \& Stark, G. R. [1999]; \textit{Mechanisms of G2 arrest in
response to overexpression of p53}, Mol. Biol. of the Cell 10, 3607-3622.

Thieffry, D. \& Romero, D. [1999]; \textit{The modularity of biological
regulatory networks}, BioSystems 50, 49-59.

Thomas, R., Thieffry, D. \& Kaufman, M. [1995]; \textit{Dynamical behavior
of biological regulatory networks - I. Biological role of feedback loops and
practical use of the concept of loop-characteristic state}, Bull of Math.
Biology 57, 247-276.

Vilela Mendes, R. [2003]; \textit{Tools for network dynamics},
cond-mat/0304640, to appear in Int. J. Bifurcation and Chaos.

Vogelstein, B., Lane, D. \& Levine, A. J. [2000]; \textit{Surfing the p53
network}, Nature 408, 307-310.

Volchenkov, D. \& Lima, R. [2003]; \textit{Homogeneous and scalable gene
expression regulatory networks with random layouts of switching parameters},
q-bio.MN/0311031

Vousden, K. H. [2000]; \textit{p53: Death star}, Cell 103, 691-694.

Woods, D. B. \& Vousden, K. H. [2001]; \textit{Regulation of p53 function},
Exp. Cell Research 264, 56-66.

\end{document}